# How Fast You Can Actually Fly: A Comparative Investigation of Flight Airborne Time in China and the U.S.

Ke Liu[1], Zhe Zheng[2,3*], Bo Zou[2], Mark Hansen[1]

[1] Department of Civil and Environmental Engineering, University of California, Berkeley, United States
[2] Department of Civil and Materials Engineering, University of Illinois at Chicago, United States
[3] National Key Laboratory of Air Traffic Flow Management, Nanjing University of Aeronautics and Astronautics, China

January 17, 2020

**Abstract:** Actual airborne time (AAT) is the time between wheels-off and wheels-on of a flight. Understanding the behavior of AAT is increasingly important given the ever growing demand for air travel and flight delays becoming more rampant. As no research on AAT exists, this paper performs the first empirical analysis of AAT behavior, comparatively for the U.S. and China. The focus is on how AAT is affected by scheduled block time (SBT), origin-destination (OD) distance, and the possible pressure to reduce AAT from other parts of flight operations. Multiple econometric models are developed. The estimation results show that in both countries AAT is highly correlated with SBT and OD distance. Flights in the U.S. are faster than in China. On the other hand, facing ground delay prior to takeoff, a flight has limited capability to speed up. The pressure from short turnaround time after landing to reduce AAT is immaterial. Sensitivity analysis of AAT to flight length and aircraft utilization is further conducted. Given the more abundant airspace, flexible routing networks, and efficient ATFM procedures, a counterfactual that the AAT behavior in the U.S. were adopted in China is examined. We find that by doing so significant efficiency gains could be achieved in the Chinese air traffic system. On average, 11.8 minutes of AAT per flight would be saved, coming from both reduction in SBT and reduction in AAT relative to the new SBT. Systemwide fuel saving would amount to over 300 million gallons with direct airline operating cost saving of nearly $1.3 billion nationwide in 2016.

---

[*] Ke Liu and Zhe Zheng are both first authors.

# 1 Introduction

With the total number of global air travelers growing tenfold from 1970 to 2017 (World Bank, 2018), air traffic systems worldwide have become increasingly complicated and in many situations congested, causing negative consequences to airlines and the traveling public such as increased airline operating cost, loss of passenger welfare, and added fuel consumption and emissions (see, e.g., Ball, et al., 2010; Zou and Hansen, 2012; Dresner et al., 2012; Ryerson et al., 2014). To mitigate these negative consequences, the airline industry and air traffic control (ATC) have made significant efforts. For example, it is common practice for airlines to pad extra time into flight schedules to absorb and prevent delay from being propagated. Integrated disruption management with flight planning has also been considered to help with schedule recovery once delay occurs (e.g., Clausen et al., 2010; Marla et al., 2016). On the ATC side, many traffic management procedures including ground delay programs (GDP), airspace flow programs (AFP), and Area Navigation (RNAV) have been implemented to balance flight demand with terminal and en-route airspace capacity and stream air traffic. The goal of these efforts is to minimize delay and maximize the extent to which aircraft fly according to schedule. To do so, understanding the behavior of actual airborne time (AAT), i.e,. the time between the actual wheels-off at the departure airport and the actual wheels-on at the arrival airport which is the most expensive part of a flight, is of critical importance.

The behavior of AAT is intertwined with other time components related to a flight. As shown in Figure 1, AAT is part of the actual block time (ABT) of a flight which consists of AAT and taxi-out and taxi-in times. For a typical flight, taxi times hold a small proportion compared to AAT suggesting a high correlation between AAT and ABT. In an ideal situation, ABT exactly coincides with scheduled block time (SBT) of a flight. However, in reality ABT often deviates from SBT due to departure and arrial delays. The discrepancy between AAT and SBT thus stems from departure delay, taxi-out time, arrival delay, and taxi-in time, the former two forming the ground time at the departure airport (D_GT).

After arrival at gate, the extent of after-flight turnaround time (AF_TT) plus arrival delay relative to the needed time for turnaround determines whether delay will incur to the gate departure of the next flight.

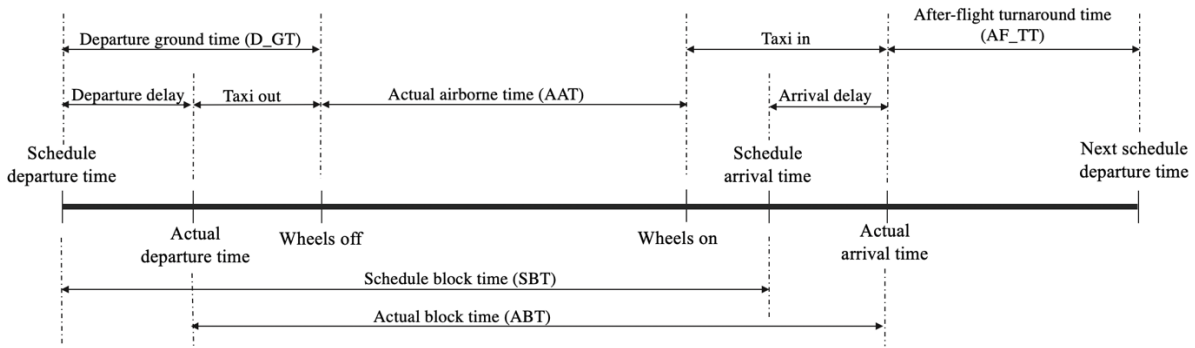

Figure 1. Main decomposition of air travel time

The time components involved in a flight described above provide some useful perspectives on what to consider when modeling AAT. A first important aspect is the extent to which AAT is in line with SBT, which in turn is influenced by historic ABT of flights operating on the same OD. A closely related issue is how AAT is aligned with the physical distance flown. Second, as on-time performance is a main concern of flying, an important factor to consider is the amount of time a flight spends prior to takeoff and whether pre-takeoff delay leads a flight to speed up to reduce delay. Similarly, the amount of turnaround time between the actual gate arrival of a flight and the schedueld gate departure of the next flight may affect AAT of the flight. Anticipating a small turnaround time, the ability of a flight to shorten AAT to reduce arrival delay should be explored. Apart from the above, other flight operation-related factors such as aircraft heading, hub status of departure/arrival airports, and the position of the flight in a day's schedule are also expected to exert some influence on AAT. AAT are likely to be further influenced by time-of-a-day, day-of-a-week, seasonal, and weather-related factors.

To our knowledge, no research has looked into the behavior of AAT. Even the efforts to model ABT and SBT are limited. For ABT, Coy (2006) employed a two-stage statistical model with one-year flight records to estimate the block time of U.S. commercial flights. Factors chosen for ABT estimation include historical ABT, flight arrival time, airport utilization, ice, and the interaction of poor weather conditions with traffic. Deshpande and Arikan (2012) used Ordinary Least Square (OLS) to estimate truncated block time, defined as ABT minus late aircraft delay, as a function of route, carrier, airports, airport congestion and aircraft-specific variables. In a series of studies, the question of how the variety of ABT distribution affects airline SBT setting considering flight on-time performance has been examined (Mayer and Sinai, 2003; Hao and Hansen, 2014; Kang and Hansen, 2017, 2018). More recently, Fan (2019) investigated the variation of SBT over the past three decades using quantile and OLS regressions with air traffic condition, flight delays, aircraft type, flight heading, airports slot policy, airport-specific anomalies, airline-specific policies, and changes in crude oil price as explanatory variables. It is found that worldwide, SBT has grown by 0.21-0.33 minutes per year.

In view of the existing literature, this paper aims to make two major contributions. The first one is the explicit modeling of AAT and testing of its relationship with a range of operation-, time-, and weather-related factors. In particular, we are interested in the extent to which AAT sticks to SBT and its correlation with flight distance. Also of interest to us is the interactions of AAT with pre-takeoff time and after-landing turnaround time, given that the three time components together determine delay propagation downstream. A large amount of time spent prior to takeoff may pressure a flight to reduce AAT to catch up with the schedule. Knowning that the turnaround time is small after landing, a flight may also try to reduce AAT to prevent arrival delay and its passing to connected operations. These conjectures, which are unexplored previously, are part of the hypotheses to be tested in our study. Furthermore, how the AAT behavior evolves as a flight flies longer and as an aircraft flies more legs in a day will be explored as well through sensitivity analysis.

For the second major contribution, our investigation of AAT behavior is conducted in a comparative manner for flights in the U.S. and China. The air traffic systems of the two countries are the two largest in the world. The two countries are also similar in territory size, both located in the belts of prevailing westerlies and dealing with significant flight delays (Ren and Li, 2018). Yet significant differences exist in their air traffic systems in terms of geographical distributions of air travel demand, airspace availability, air traffic flow management, delay management policies by the respective civil aviation authorities, and airline scheduling practices. For the geographical distributions of air travel demand, most of the economic centers in China are concentrated in the eastern part of the country, resulting in majority of flights being short- and medium-haul. In contrast, the economic centers in the U.S. are more evenly scattered throughout the country, leading to longer flight lengths (Figure 2). In terms of airspace availability, only 20% of the national airspace in China is made available for civil aviation use while in the U.S. the percentage is 80%. This stark discrepancy means more limited airspace capacity and flight routing flexibility in China (Hsu, 2014). These differences make the comparison between the two countries worthwhile, with the results possibly inform future flight scheduling and operation strategies in China to improve its system efficiency.

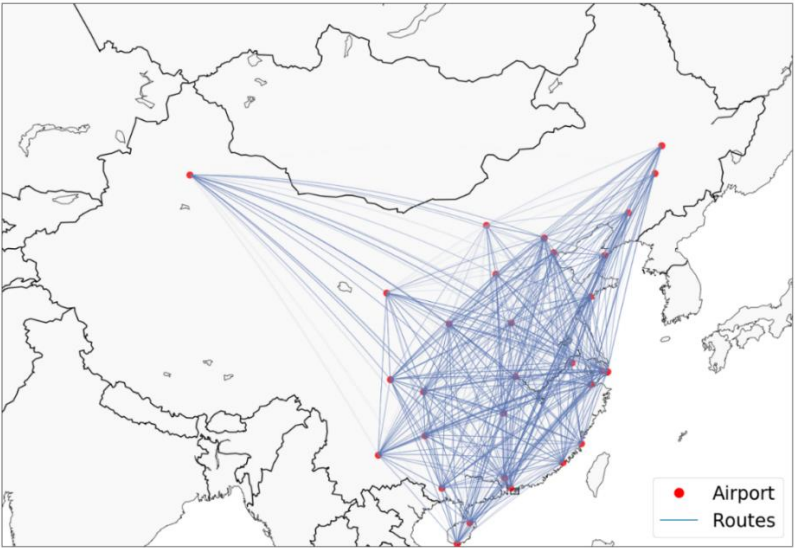

(a) Routes among top 30 airports in China

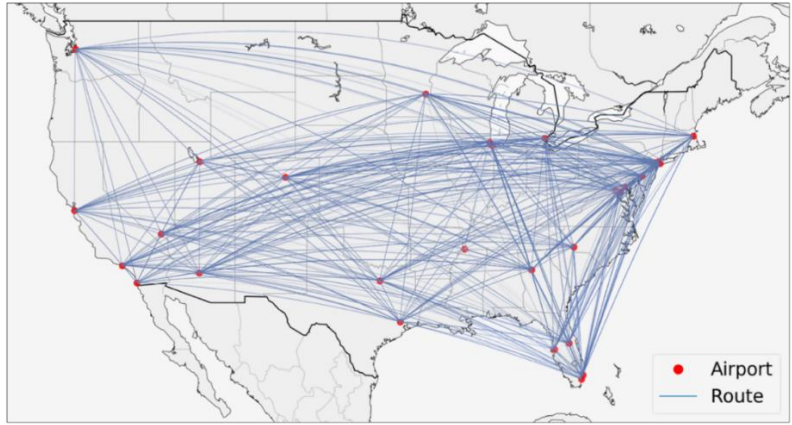

(b) Routes among top 30 airports in the U.S.

Figure 2. Air routes among top 30 airports (based on total flight operations in 2016)

in China and the U.S.

The remainder of the paper is organized as follows. In Section 2, we present four key hypotheses, which form the basis of our model specifications. Section 3 describes the data sources and the data filtering process. Estimation results of the base model are reported and discussed in Section 4. Further sensitivity analysis of the estimation results with respect to flight length and aircraft utilization is performed in Section 5. As we conjecture a more efficient flying environment in the U.S., Section 6 conducts a counterfactual analysis to assess the potential efficiency gains if the Chinese air traffic system adopted the AAT behavior in the U.S. Findings of the paper are summarized in Section 7 along with suggestions for future research.

## 2 Model development

Based on the literature review and conjecture in Section 1, in this section we identify several factors that may influence AAT. Four key hypotheses about the factor effects are formalized in subsection 2.1. Building on the hypotheses, multiple model specifications are proposed in subsection 2.2 to characterize the AAT behavior.

## 2.1 Hypotheses

AAT is closely related to scheduled time of a flight. While scheduled airborne time is not commonly reported, SBT has been widely used in aviation research and practice. Compared to AAT which involves only airborne time, SBT encompasses time both in the air and on the ground. Based on Figure 1, the relationship between AAT and SBT can be established as Eqs. (1)-(2):

$$ABT = AAT + \text{Taxi-in time} + \text{Taxi-out time} \tag{1}$$
$$SBT = ABT + \text{Departure delay} - \text{Arrival delay} \tag{2}$$

Eq. (1) indicates that ABT is the sum of AAT, actual taxi-in time, and actual taxi-out time. In Eq. (2), the difference between SBT and ABT is the difference of departure and arrival delays of the flight. Substituting ABT in Eq. (2) by its expression of Eq. (1) we obtain:

$$SBT = AAT + \text{Departure delay} - \text{Arrival delay} + \text{Taxi-in time} + \text{Taxi-out time} \tag{3}$$

In most cases that arrival delay of a flight is not too large compared to departure delay (with the difference less than the sum of taxi-in and taxi-out times), SBT will be greater than AAT. The difference between depature and arrival delays are attributed to two main factors. First is airborne delay. Second is flight buffer, i.e., the extra time padded into flight schedule to absorb delay which has become common practice in airline SBT setting. Given flight buffer and as long as no significant airborne delay occurs, arrival delay is likely to be smaller than departure delay. Further adding taxi times suggests a high likelihood that $SBT - AAT > 0$.

On the other hand, almost always, delay and taxi times are much shorter than AAT and SBT. The relativity of delay and taxi times becomes smaller as flight length increases. This fact together with Eq. (3) means that SBT is likely to highly correlate with AAT. Based on the conjecture, the first hypothesis to be tested in the paper is formalized as follows:

**Hypothesis 1:** AAT is highly correlated with SBT. The ratio of AAT over SBT should be, on average, less than one. The value of the ratio for a specific flight varies with SBT setting strategy, actual operational conditions, and flight length.

Arrival delay of a flight not only increases operating cost of the flight, but can propagate to downstream flights of the same aircraft and to different connected aircraft through crew rotation and passenger connections (Churchill et al., 2010; Kafle and Zou, 2016). To mitigate delay propagation, a flight delayed at departure may speed up en route or takes a shorter air route to reduce airborne time (Delgado, et al. 2013). Provided that there is room for reducing airborne time, the larger the pre-takeoff delay, the larger pressure a flight will have to reduce AAT. This leads to the second hypothesis of the paper.

**Hypothesis 2:** Pre-takeoff delay pressures a flight to reduce AAT. As such, AAT would be shorter with larger pre-takeoff delay.

The third hypothesis of the paper also relates to delay propagation but from a different perspective. If the turnaround time, defined as the time between the scheduled arrival of the current flight and the scheduled departure of the next flight, is longer, then the turnaround time has greater capacity to absorb and prevent any arrival delay from being propagated to the next flight. On the other hand, turnaround time should not be exceedingly long as it affects the number of legs an aircraft can fly in a day and thus airline productivity. Nonetheless, a shorter turnaround time at the destination airport may pressure a flight to fly faster to minimize arrival delay. This leads to the third hypothesis:

**Hypothesis 3:** Flights with smaller turnaround time will have more urgency to reduce AAT, in order to minimize arrival delay.

Finally, as our study attempts to compare the AAT behavior in China and the U.S., recognizing the operational differences between the two countries is important. It is conjectured that the U.S. has a more efficient air traffic system due to higher airspace availability and a more flexible routing network which allow aircrafts to adjust speed and route. Air traffic flow management (ATFM) schemes and new technologies including GDP, AFP, Performance Based Navigation (PBN), and Automatic Dependent Surveillance-Broadcast (ADS-B) also contribute to more efficient management of traffic flow en route and around terminal airspace in the U.S., thereby smoothing traffic and mitigating system congestion. This paper examines whether and to what extent the AAT behavior is different between the two countries. A counterfactual analysis is conducted to understand the implications for system-wide efficiency gains if the AAT behavior in China became the same as in the U.S. This brings the fourth hypothesis of the paper.

**Hypothesis 4:** Provided that the air traffic system in China had the same operation environment as in the U.S., Chinese flights would have overall shorter AAT leading to operating cost and fuel savings.

## 2.2 Model specifications

Building on the first three hypotheses, this subsection specifies econometric models for AAT to be estimated in the paper. The general form of the AAT models is presented in Eq. (4), in which explanatory variables affecting AAT are classified into three categories: operation related, time related, and weather related. The specific variables considered in the model specifications are listed in Table 1.

$$AAT = f(\text{operation-related variables, time-related variables, weather-related variables}) \quad (4)$$

For operation-related variables, as mentioned earlier the most relevant variable is SBT and the origin-destination (OD) distance, to represent which we use the great circle distance (GCD) between the origin and the destination airports. Because SBT and GCD are highly correlated, signficaint collinearity would arise if both are included in the model. To understand the impact of SBT and GCD on AAT, two separate specifications are considered: one with only SBT, termed SBT model as shown in Eq. (5); the other with only GCD, termed GCD model as shown in Eq. (6). For the GCD model, we further include a variable SBT_dif, which is the difference between SBT of the flight and the average SBT of all flights of the same OD, airline, aircraft type, and quarter. As the GCD model does not have other schedule related variables, the SBT_dif variable is intended to capture the incentive to fly faster/slower suggested by the schedule relative to the average schedule.

Other than SBT and GCD, the SBT and the GCD models share a number of common operation-related variables. One variable is D_GT, which is the the ground time at the departure airport, or the actual wheels-off time minus the scheduled gate departure time. The D_GT variable is used as a substitute for gate departure delay given that the Chinese data do not provide information on the actual gate departure time. Recall from Figure 1 that D_GT includes both gate departure delay and taxi-out time. Thus, the D_GT variable also captures the effect on AAT of taxi-out delay. This variable is used to test Hypothesis 2.

To test Hypothesis 3, a variable AF_TT_dif is constructed as the difference between the the after-flight turnaround time (AF_TT) and the minimum connection time (MCT) at the destination airport. AF_TT is calculated as the difference between scheduled departure time of the next flight of the same aircraft and scheduled arrival time of the current flight under study. In China, the Civil Aviation Administration of China (CAAC, 2012) documents MCT values but does not provide details on how the values are calculated. In the U.S., such information is not available from the Federal Aviation Administration (FAA). To make the measurement of MCT consistent in the two countries, we use $5^{th}$ percentile value of the actual turnaround time of similar flights at the same airport. Similar flights refer to flights that have the same aircraft type, arrival airport, and airline. The use of the $5^{th}$ percentile rather than the observed minimum turnaround time follows the idea of existing work (e.g., Britto et al., 2012; Zou and Hansen, 2012) to make the calculation more robust to measurement error and reduces the influence of unusually favorable conditions. The expectation is that if a flight's AF_TT is less than MCT, then the flight may be more pressured to reduce AAT to minimize arrival delay.

We consider a Heading variable, which measures the longitude difference between the departure and the arrival airports, with eastward taking positive values and westward taking negative values. The expectation is that prevailing winds may affect flying speed and thus AAT. We use HubDep and HubArr to indicate whether a flight takes off from or lands at the airline's hub. We conjecture that a flight from/to a hub airport is likely to receive better ground services. In addition, an airline usually has a good relationship with ATC at its hub airports, which further improves flight operations. Another binary variable indicating whether a flight is the last leg flown by an aircraft is included, to test whether being the last flight in a day will lead to faster or slower flying.

Time-related factors are expected to affect AAT as well. Several variables are considered in this category. The first one is DT, a binary variable indicating whether a flight operates in the day or night. DT equals 1 if the scheduled departure time is between 6:00 and 20:00. The hypothesis is that AAT may be higher in the day due to greater traffic and congestion. Second, AAT may vary depending on it is a weekday or in the weekend, again considering the difference in air traffic particularly for business travel. A Wkd variable is included which takes value 1 if scheduled departure day is a weekday and 0 otherwise. Third, traffic pattern differences arising from seasonal changes in air travel demand and consequently airline scheduling strategies may further affect AAT. To this end, three seasonable variables—Sprg, Sumr and Fall—are introduced.

To capture the impacts of weather on AAT, we consider temperature, wind speed, and visibility at departure and arrival airports of a flight. For temperature, we postulate that both hot and cold temperatures present adverse weather and impact the AAT negatively. Thus the absolute difference between the temperature at the actual departure / arrival time of a flight and the average temperature associated with all departure / arrival flights at the same airport is considered (TempDepDif and TempArrDif). The wind speed variable, WdspDep / WdspArr, provides the speed of wind at the departure / arrival airport during the hour of the actual departure / arrival time of the flight. The visibility variables VisDep / VisArr are similarly defined but taking logarithmic values to capture the plausible nonlinear relatiohship between visibility and AAT. For example, visibility decreasing from 1,000 meters to 500 meters impacts flight operations more than decreasing from 8,000 meters to 7,500 meters. To the extent that some records have visibility of zero miles, we add one to the exponent: $VisDep = \ln(\text{Departure Visibility} + 1)$ and $VisArr = \ln(\text{Arrival Visibility} + 1)$.

Following the discussions above, below are the full specification of the SBT and GCD model specifications. In the specifications, $\alpha$'s and $\beta$'s are the coefficients to be estimated. $\varepsilon$ and $\upsilon$ are the error terms.

$$\begin{aligned} AAT = &\alpha_0 + \alpha_1 SBT + \alpha_2 D\_GT + \alpha_3 AF\_TT\_dif + \alpha_4 Heading + \alpha_5 HubDep + \alpha_6 HubArr \\ &+ \alpha_7 Flt\_last + \alpha_8 DT + \alpha_9 Wkd + \alpha_{10} Sprg + \alpha_{11} Sumr + \alpha_{12} Fall + \alpha_{13} TempDep \\ &+ \alpha_{14} TemArrDif + \alpha_{15} WdspDep + \alpha_{16} WdspArr + \alpha_{17} VisDep + \alpha_{18} VisArr + \varepsilon \end{aligned} \quad (5)$$

$$\begin{aligned} AAT = &\beta_0 + \beta_1 GCD + \beta_2 SBT\_dif + \beta_3 D\_GT + \beta_4 AF\_TT\_dif + \beta_5 Heading + \beta_6 HubDe \\ &+ \beta_7 HubArr + \alpha_8 Flt\_last + \beta_9 DT + \beta_{10} Wkd + \beta_{11} Sprg + \beta_{12} Sumr + \beta_{13} Fall \\ &+ \beta_{14} TempDepDif + \beta_{15} TemArrDif + \beta_{16} WdspDep + \beta_{17} WdspArr + \beta_{18} VisDep \\ &+ \beta_{19} VisArr + \upsilon \end{aligned} \quad (6)$$

Table 1. Description of variables

| Variables | | Description |
|---|---|---|
| *Y-variables* | | |
| | AAT (min) | Actual airborne time, which is the time difference between actual wheels-on time and actual wheels-off time |
| *X-variables* | | |
| **Operation related** | SBT (min) | Scheduled block time |
| | GCD (mile) | Great circle distance between departure and arrival airports |
| | SBT_dif (min) | Difference between SBT and the average SBT of all flights of the same OD, airline, aircraft type, and quarter |
| | D_GT (min) | Gound time at the departure airport, calculated as: Actual wheels-off time – Scheduled gate departure time |
| | AF_TT (min) | Scheduled turnaround time, calculated as: Scheduled gate departure time of the next flight – scheduled gate arrival time of the current flight |
| | AF_TT_dif (min) | Difference between AF_TT and the minimum connection time (MCT): AF_TT – MCT |
| | Heading (degree) | Longitude difference between departure and arrival airports, positive if eastward and negative if westward |
| | HubDep / HubArr | Takes value 1 if the flight departs from / arrives at a hub airport |
| | Flt_last | Takes value 1 if the flight is the last leg flown by an aircraft in a day |
| **Time related** | DT | Takes value 1 if the flight operates in the day, i.e., scheduled departure between 6:00 and 20:00, and 0 otherwise |
| | Wkd | Takes value 1 if the flight operates on a weekday, and 0 otherwise |
| | Sprg / Sumr / Fall | Takes value 1 if the flight is scheduled to depart in Spring / Summer / Fall, and 0 otherwise |
| **Weather related** | TempDepDif / TempArrDif (Fahrenheit) | Absolute difference between the temperature at the actual departure / arrival time of a flight and the average temperature associated with all departure / arrival flights at the same airport |
| | WdspDep / WdspArr (Knots) | Windspeed at the actual departure / arrival time of a flight at the corresponding airport |
| | VisDep / VisArr | Log of visibility (in miles) at the actual departure / arrival time at of a flight at the corresponding airport |

# 3 Data

## 3.1 Data sources

The data needed for model estimation come from several sources. For operation- and time-related variables, flight-level records are collected for the year 2016 from the Air Traffic Management Bureau of CAAC on the China side and the the Aviation System Performance Metrics (ASPM) of FAA on the U.S. side. This results in 3,705,093 and 6,883,674 flight records collected in China and the U.S. respectively. On the China side, the flight records include commercial and general aviation flights. On the U.S. side, the flight records consist of commercial, general aviation, air taxi, and military flights. Each record documents the departure and arrival airports, departure and arrival dates, airline to which the flight belongs, aircraft type of the flight, tail number of the aircraft, the scheduled gate departure time, the actual wheels-off time, the actual wheels-on time, and the scheduled gate arrival time. The time information allows us to construct AAT, SBT, SBT_dif, D_GT, AF_TT, AF_TT_dif, and Flt_last, DT, Wkd, and the seasonal dummy variables. In addition, GCD between any two airports is calculated based on the latitudes and longitudes of the airports, the information of which is collected from OpenFlight (2019). The longitude information also allows for construction of the Heading variable.

For weather-related variables, the Meteorological Terminal Aviation Routine (METAR) weather report provides weather information including temperature, windspeed, and visibility at each of the airports covered in the data on the China side, at 60-minute intervals. On the US side, the ASPM data offers similar weather records, at a higher frequency of every 15 minutes.

## 3.2 Data filtering

Multiple data filtering steps are performed to ensure data consistency. First, as we are interested in AAT of commercial flights, other types of flights are removed. Second, we note that the U.S. has more airports than China. Many of the airports in the U.S. are mainly for general aviation, whereas there are very few general aviation airports in China. To ensure that the AAT comparison is meaningful, we only keep flights that flew between the top 30 airports in terms of total flight operations in each country in 2016. These airports are shown in Figure 1 and also listed in Appendix 1. This filtering reduces the number of flight records to 1,815,416 for China and 2,465,472 in the U.S.

In addition to the above, records with missing or abnormal variable values are eliminated from the data. Flights with GCD less than five miles between the departure and the arrival airports are deleted, as it might indicate that a flight flew between two airports of the same city (e.g., Hongqiao and Pudong airports in Shanghai). If GCD is zero miles, it might also suggest that the flight was canceled. For each OD pair, we delete flight records with AAT outside 1.5 times the difference between the $85^{th}$ and the $15^{th}$ percentile values in the AAT distribution, above the $85^{th}$ percentile and below the $15^{th}$ percentile. These flights are viewed as outliers as they have exceedingly long or short AAT. Similar filtering of flight records with extremely long and short D_GT is also performed. After the filtering, the China dataset has 1,723,853 flights while the U.S. dataset has 2,357,067 records. The summary statistics for some of the variables are reported in Table 2.

Table 2. Summary statistics of the variables in the specified models after data cleaning

| Variable | Countries | Mean | Std. Dev. | Min. | Max. |
|---|---|---|---|---|---|
| AAT | China | 120 | 44 | 8 | 412 |
|  | U.S. | 138 | 76 | 6 | 569 |
| GCD | China | 703 | 324 | 28 | 2,293 |
|  | U.S. | 1,038 | 650 | 53 | 2,749 |
| SBT_dif | China | 0.002 | 3.9 | -104 | 577 |
|  | U.S. | 0.019 | 4.8 | -136 | 528 |
| D_GT | China | 27 | 33 | -308 | 1153 |
|  | U.S. | 25 | 24 | -46 | 152 |
| AF_TT | China | 300 | 573 | 0 | 5,760 |
|  | U.S. | 897 | 867 | 2 | 5,678 |
| AF_TT_dif | China | 372 | 5,733 | -4,830 | 527,429 |
|  | U.S. | 1,271 | 8,062 | -3,225 | 520,225 |
| Heading | China | -0.009 | 9.2 | -27 | 27 |
|  | U.S. | -0.002 | 20.3 | -51 | 51 |
| TempDepDif | China | 15.8 | 11.3 | 0.5 | 93.1 |
|  | U.S. | 14.7 | 10.6 | 0.2 | 82.2 |
| TempArrDif | China | 15.7 | 11.2 | 0.4 | 93.2 |
|  | U.S. | 14.7 | 10.6 | 0.2 | 82.2 |
| WdspDep | China | 6.5 | 4.1 | 0 | 35.0 |

|        | U.S.  | 9.1 | 2.7 | 1 | 21.8 |
|--------|-------|-----|-----|---|------|
| WdspArr | China | 6.5 | 4.1 | 0 | 33.0 |
|        | U.S.  | 9.3 | 2.7 | 1 | 21.8 |
| VisDep | China | 1.6 | 0.4 | 0 | 1.9 |
|        | U.S.  | 2.2 | 0.3 | 0 | 2.9 |
| VisArr | China | 1.6 | 0.4 | 0 | 1.9 |
|        | U.S.  | 2.2 | 0.3 | 0 | 2.9 |

# 4 Estimation results

This section presents the estimation results of the AAT models specified in Section 2. The GCD model is estimated using OLS. For the SBT model, endogeneity may arise as it is expected that SBT is set based on AAT. Hausman test is conducted on the explanatoary variables in the SBT model. The test result rejects the null hypothesis that the OLS estimator is consistent. Therefore, instrumental variables are needed in place of SBT. In this study, we consider GCD, dummies indicating the hours of scheduled departure and arrival, airline, and aircraft type, and averaged weather conditions as additional instruments related to SBT setting. The averaged weather conditions include temperature, wind speed, and visibility of the departure and arrival hours (at the departure and arrival airports), with the average over all days of the same month. As historic weather information before 2016 is not available to us, the average weather conditions of the current month is used as a proxy for the weather conditions in the preceding years of the month, which airlines would use when setting SBT. These additional instruments are considered exgeonous to AAT. With these added instrumental variables, we estimate the SBT model using generalized method of moments (GMM), which is more efficient than two-stage least square estimation (Green, 2003).

The estimation results are presented in Table 3. Model 1 considers SBT and is estimated using GMM. Model 2 has the same specification as Model 1 but is estimated using OLS, serving as a point of reference. Model 3 considers GCD and SBT_dif in place of SBT. GCD is clearly exogenous. For SBT_dif, one may suspect that the difference of a flight's SBT from the average may be affected by AAT. However, finding adequate instruments for SBT_dif is difficult. We have experimented with using the instruments of Model 1, but they turn out to have very low correlation with SBT_dif ($R^2$ of regressing SBT_dif on the instruments is below 0.05 for both countries). Moreover, whether considering SBT_dif as endogenous or exogenous, the estimates for other coefficients than SBT_dif do not seem to be quite different. In view of the above, we still report OLS estimation results for the GCD model in Table 3. All standard errors are clustered by OD pair.

Looking at the individual coefficients, in Model 1 the SBT coefficients are 0.97 for both China and the U.S., which confirms Hypothesis 1 that a high correlation exists between AAT and SBT. In addition, the less-than-one value indicates that the flight buffer effect overweighs the airborne delay effect resulting in smaller AAT than SBT. It is important to keep in mind that the same coefficients for China and the U.S. only suggest the same relationship between AAT and SBT. They do not mean that aircraft in the two countries fly the same speed. Rather, the difference in flying speed is reflected in the scheduling practices which lead to different SBT for ODs that have similar characteristics in the two countries.

Model 3 provides further evidence on the difference in flying speed in the two countries. The GCD coefficients, which are 0.124 for China and 0.115 for the U.S., imply an average flying speed of 484 mph in China and 521 mph in the U.S. They are obtained by multiplying the inverse of the GCD coefficients by 60. As most of the Chinese airspace is for military use, the smaller GCD-based speed is associated with a greater extent of detour that Chinese flights have to make to avoid intruding on military airspace. For the SBT_dif coefficient, it is statistically significant for both China and the U.S. but small in magnitude (0.096 for China and 0.140 for the U.S.), suggesting that a shorter schedule would make a flight fly faster, but only slightly. For example, if a flight has one less minute in its schedule than the average, then AAT would reduce by about 0.1 minutes.

Recall that apart from SBT, GCD, and SBT_dif, the other variables are the same in Models 1 and 3. Thus, it begs the question of which model should be the focus of our interpretation. We opt for Model 1, as it has more coefficients with statistical significance and greater goodness-of-fit, especially on the China side. Moreover, by choosing SBT rather than GCD as an explanatory variable, we account for not only the distance effect, but also the effect of scheduling practices that are afftected by en-route airspace availability and air traffic control which are different in the two countries.

In Model 1, the D_GT coefficients are negative for both countries, meaning that more ground time at the departure end leads to shorter AAT. This confirms Hypothesis 2 that delay at the departure end will pressure a flight to fly faster. However, the effect is quite small: if a Chinese flight spends 20 additional minutes on the ground prior to takeoff, on average the flight could fly faster to shorten AAT by less than half a minute (0.48 minutes). The number is somewhat higher for a U.S. flight (0.68 minutes). The estimates suggest limited possibility for flights in both countries to speed up to recover pre-takeoff delays. The limited possibility may be attributed to the established ATC procedures and airspace congestion in both countries which suggest small leeway for a flight to reduce AAT through speed or route adjustment. Particualrly for the U.S., the wide use of GDP means that ground delay is often part of ATFM. As such, ground delay is not unexpected. Therefore, a flight with some ground delay does not feel imperative to speed up to catch up with the original schedule.

The next variable of our interest is AF_TT_dif, which has a positive sign for both countries and thus consistent with our *a priori* expectation: flights with smaller turnaround time will have more urgency to reduce AAT. Again, the coefficient is so small in both countries (in the order of $10^{-5}$) that it basically means that the effect is non-existent. In other words, a flight would not really consider the amount of turnaround time after landing while deciding whether to speed up or not. The null of Hypothesis 3 is rejected.

For the remaining operation-related variables, the Heading variable has significant coeffceints but of opposing signs in the two countries. As China and the U.S. are on the similar latitudes of the belts of prevailing westerlies, flights flying eastward are expected to spend less time than flights flying westward. The China coefficient is consistent with this expectation. For the U.S. model, a possible explanation for the counter-intuitive sign is that the prevailing wind effect may have been incorporated in the setting of SBT rather than AAT conditional on SBT. To explore this further, we regress SBT on all other explanatory variables and the added instrumental varaibles. We find a negative coefficient for the Heading variable for both countries.

For the hub airport variables, the negative coefficients reveal that using a hub airport either at the departure or arrival end will help reduce AAT. This is reasonable as such flights may receive better ATC support near the terminal airspace. While the HubDep effect is larger than HubArr effect in China, estimation in the U.S. is the opposite. The most signficnat AAT reduction occurs at the departure end in China: a flight departing from a hub airport will have 4.37 minutes less AAT than leaving a non-hub airport. In the U.S., the effect is only 0.59 minutes.

Finally, the Flt_last coefficient is negative with a larger magnitude in China as opposed to a positive but smaller coefficient for the U.S. Conceptually, two opposing effects may affect Flt_last. On the one hand, being in the last flight leg which is typically late in a day the crew have the motivation to get off work earlier by flying faster. On the other hand, the last leg of an aircraft means no worry about delay propagating downstream. This relieves the pressure to fly fast. The estimation results show that for China, the first effect overweighs the second effect, whereas in the U.S. the second effect dominates.

Turning now to the the time- and weather-related variables, the coefficients for DT have oppositive signs for China and the U.S. The positive sign for China is intuitive: more traffic in the day means more congestion and thus longer AAT. The negative sign in the U.S. may be interpreted from a different perspective: anticipating more daytime traffic and congestion, U.S. flights may self-motivate to fly faster. Either way, the DT effect is not very large: less than one minute in both countries. For the similar consideration that more traffic during weekdays leads to greater AAT, the Wkd variable has a positive coefficient. Flying in the winter season tends to be faster than during spring and summer, but slower than in the fall. The tendency is consistent in China and the U.S. For the weather-related variables, they generally have small effects. The most significant effect occurs to the arrival visibility (VisArr), whereas the effect of departure visblity (VisDep) is much smaller (for the U.S., insignficiant). The difference is not surprising, as low visibility will mainly affect ground holding time before takeoff, not afterwards (i.e., AAT). In contrast, better visibility at the arrival end helps decrease separation while forming the landing queue which contributes to reducing AAT.

Besides the above coefficients, the estimates for the constant term are also worth discussions. As AAT does not involve ground operation time but SBT does, the constant may be viewed as a proxy for the average taxi time, estimated to be 19.1 minutes in China and 21.3 minutes in the U.S. Furthermore, note that in Model 3 GCD is used instead of SBT. The meaning of the constant in Model 3 is thus different, reflecting the additional time a flight spends on acceleration and decelartion while airborne. The estimates are 23.6 minutes for Chinese flights and 22.1 minutes for U.S. flights.

A last point to notice is $R^2$'s of the models. Across Models 1-3, the U.S. side has a greater goodness-of-fit than the China side. The larager $R^2$'s on the U.S. side can be attributed to the wide use of GDP which intends to substitute as much airborne delay as possible by ground delay. In contrast, GDP was not implemented in China in 2016. Another possible reason is associated with the more abundant airspace available for commercial flights in the U.S. which allows for greater probability of flying GCD routes.

Table 3. Base models for AAT behavior in China and the U.S.

| | Model 1 (GMM for SBT) | | | | Model 2 (OLS for SBT) | | | | Model 3 (OLS for GCD) | | | |
|---|---|---|---|---|---|---|---|---|---|---|---|---|
| | China | | U.S. | | China | | U.S. | | China | | U.S. | |
| | Est. | Std. err. | Est. | Std. err. | Est. | Std. err. | Est. | Std. err. | Est. | Std. err. | Est. | Std. err. |
| SBT | 0.970*** | 0.003 | 0.970*** | 0.001 | 0.950*** | 0.006 | 0.964*** | 0.002 | - | - | - | - |
| SBT_dif | - | - | - | - | - | - | - | - | 0.096*** | 0.011 | 0.140*** | 0.006 |
| GCD | - | - | - | - | - | - | - | - | 0.124*** | 0.003 | 0.115*** | 0.0003 |
| D_GT | -0.024*** | 0.001 | -0.034*** | 0.001 | -0.023*** | 0.003 | -0.037*** | 0.002 | 0.029*** | 0.006 | -0.005*** | 0.002 |
| AF_TT_dif | 5.3E-5*** | 2.2E-6 | 3.7E-5*** | 2.1E-06 | 5.3E-5*** | 1.1E-05 | 4.2E-04*** | 6.8E-05 | -1.4E-05 | 1.1E-05 | 3.08E-06 | 3.2E-06 |
| Heading | -0.085*** | 0.014 | 0.052*** | 0.005 | -0.111*** | 0.034 | 0.053*** | 0.008 | -0.276*** | 0.081 | -0.494*** | 0.008 |
| HubDep | -4.370*** | 0.211 | -0.585*** | 0.162 | -4.346*** | 0.415 | -0.179 | 0.345 | -0.354 | 0.875 | -2.214*** | 0.264 |
| HubArr | -1.118*** | 0.211 | -2.700*** | 0.170 | -0.893** | 0.424 | -2.420*** | 0.329 | -0.200 | 1.017 | -0.829*** | 0.258 |
| Flt_last | -1.266*** | 0.122 | 0.239*** | 0.067 | -0.843*** | 0.215 | 0.305** | 0.126 | 1.432*** | 0.556 | -0.075 | 0.109 |
| DT | 0.756*** | 0.143 | -0.532*** | 0.099 | 0.898*** | 0.287 | -0.655*** | 0.149 | 3.324*** | 0.517 | 1.702*** | 0.122 |
| Wkd | 0.325*** | 0.022 | 0.288*** | 0.021 | 0.340*** | 0.036 | 0.267*** | 0.045 | 0.285*** | 0.038 | 0.546*** | 0.031 |
| Sprg | 1.443*** | 0.121 | 0.617*** | 0.074 | 1.586*** | 0.200 | 0.569*** | 0.143 | 0.731*** | 0.253 | -0.069 | 0.232 |
| Sumr | 1.020*** | 0.108 | 0.726*** | 0.079 | 1.087*** | 0.194 | 0.635*** | 0.122 | -2.163*** | 0.268 | -1.849*** | 0.386 |
| Fall | -0.232* | 0.121 | -0.206*** | 0.076 | -0.267 | 0.235 | -0.280** | 0.132 | -3.346*** | 0.337 | -1.496*** | 0.299 |
| VisDep | 0.070 | 0.124 | 0.118** | 0.052 | -0.073 | 0.197 | 0.086* | 0.103 | 1.646*** | 0.484 | -0.307 | 0.127 |
| TempDepDif | 0.042*** | 0.005 | -0.019*** | 0.004 | 0.042*** | 0.009 | -0.015* | 0.008 | 0.050** | 0.024 | 0.040*** | 0.010 |
| WdspDep | -0.100*** | 0.016 | -0.068*** | 0.003 | -0.108*** | 0.031 | -0.075*** | 0.012 | 0.382*** | 0.068 | 0.075*** | 0.116 |
| VisArr | -0.711*** | 0.151 | -1.746*** | 0.059 | -0.578** | 0.240 | -1.674*** | 0.123 | -0.368 | 0.510 | -2.023*** | 0.162 |
| TempArrDif | 0.031*** | 0.007 | -0.021*** | 0.005 | 0.029*** | 0.013 | -0.029*** | 0.009 | 0.023 | 0.021 | -0.001 | 0.013 |
| WdspArr | 0.129*** | 0.015 | 0.064*** | 0.003 | 0.128*** | 0.032 | 0.061*** | 0.012 | 0.153*** | 0.058 | 0.109*** | 0.012 |
| Cons | -19.108*** | 0.474 | -21.27*** | 0.324 | -16.836*** | 1.022 | -20.646 | 0.662 | 23.591*** | 2.221 | 22.068*** | 0.753 |
| Num. Obs | 1,723,853 | | 2,357,067 | | 1,723,853 | | 2,357,067 | | 1,723,853 | | 2,357,067 | |
| R-sq | 0.935 | | 0.981 | | 0.936 | | 0.982 | | 0.858 | | 0.983 | |

*** $p<0.01$, ** $p<0.05$, * $p<0.10$.

# 5 Sensitivity analysis

This section investigates the sensitivity of the estimation results to flight length and aircraft utilization. For flight length, recall the discussion of Hypothesis 1 that the difference between SBT and AAT is a function of departure and arrival delays and taxi times. The longer the flight length, the smaller the relativity of delays and taxi times with respect to AAT and SBT. Longer flight length also means greater interaction with the en-route airspace environment, which differs significantly in China and the US. For aircraft utilization, its impact on AAT is shaped by two opposing forces: 1) tendency to add more flight buffer given the increased propensity for delay propagation from an earlier flight to a later flight. This will increase the difference between AAT and SBT; 2) pressure to reduce flight buffer to be able to fly more flights. The investigation of AAT by aircraft utilitization thus helps understand which force dominates in each of the two countries.

## 5.1 Sensitivity to flight length

We divide the flight records in each country into three flight length groups based on GCD values: 1) less than 800 nautical miles as short-haul flights; 2) between 800 and 1,600 nautical miles as medium-haul flights; and 3) greater than 1,600 nautical miles as long-haul flights. Figure 3 shows the number of flights in each group in the two countries. As the top 30 airports in China are concentrated in the eastern part of the country (see Figure 1), most flights are short-haul. The busiest routes, notably Beijing-Shanghai Pudong (PEK-PVG), Shanghai Hongqiao-Shenzhen (SHA-SZX), and Beijing-Guangzhou (PEK-CAN), all fall into the short-haul category. Long-haul flights hold a very small portion in total. In contrast, the top 30 airports in the U.S. are dispersed across the nation, flights are more evenly distributed among the three flight groups. Among the busiest routes, New York (JFK)-Los Angeles (LAX) is a long-haul route, Los Angeles (LAX)-San Francisco (SFO) is a short-haul route, and Los Angeles (LGA)-Chicago (ORD) is a medium-haul route. For each flight group in each country, a separate SBT model is estimated using GMM. The results are reported in Table 4. All standard errors are clustered by OD pair.

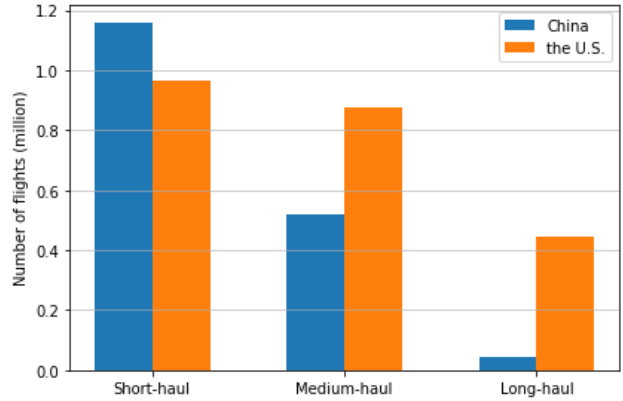

Figure 3. Flight distribution by flight length in China and the U.S.

In line with the hypotheses in Section 2, we focus on comparing the SBT and D_GT coefficients in these models. AF_TT_dif coefficients are very small and effectively zero, as in the base model estimates. Figure 4 plots the point estimates of SBT and D_GT coefficients for short-, medium-, and long-haul flights in the two countries along with the 95% confidence intervals. Figure 4a shows that the point estimates of the SBT coefficients increase with flight length both in China and the U.S., and the increase is more prominent in China. Except for long-haul flights in China, all the SBT coefficients are still less than one. These coefficients indicate that as a flight flies longer, the flight buffer effect still overweighs the airborne delay effect, but to a smaller extent relative to SBT. This might be attributed to the fact that flight buffer increases less than linearly with repect to flight range and a longer-distance flight has greater propensity to encounter airborne delays. For long-haul flights in China, the poin estimate of the SBT coeffieint is about 1.04, suggesting that SBT is not even enough to cover AAT. However, the estimate has a very large confidence interval and the statistical test fails to reject the null hypothesis that the SBT coeffient is equal to one, i.e., strict proportional change of AAT and SBT. In addition, the large confidence interval of the SBT coefficient for Chinese long-haul flights implies the greater uncertainty in AAT due to more exposure to the restricted and complex en-route environment. The increase in the SBT coefficient in the U.S., on the other hand, is much more moderate, from 0.916 to 0.950 along with consistent confidence intervals. This confirms more flexible en-route network in the U.S. airspace.

Figure 4b shows that the changing trends of the D_GT coefficient for China and the U.S. are in opposite directions. In China, the D_GT coefficient value increases with flight length. For short- and medium-haul flights, the D_GT coefficient is negative which is consistent with the base model estimates in Section 4. For long-haul flights, the point estimate of the coefficient becomes positive. However, we cannot reject the null hypothesis of a zero value. In other words, for long-haul flights departure delay may not make a flight fly faster. One explanation, same as in the previous paragraph, is the greater exposure to the restricted and complex en-route environment which prevents a flight from reducing AAT to make up for the pre-takeoff time loss. Different from that, the D_GT coefficient on the U.S. side is always negative with the absolute value slightly increasing with flight length. A possible explanation is that longer flying allows a flight to take advange of the more flexible flying environment in the U.S. by speeding up and / or rerouting to recover departure delays. Overall, we conclude that it is the distinct nature of the en-route flying environment in China and the U.S. that results in the different changing trends of the coefficients for SBT and D_GT as a flight flies longer.

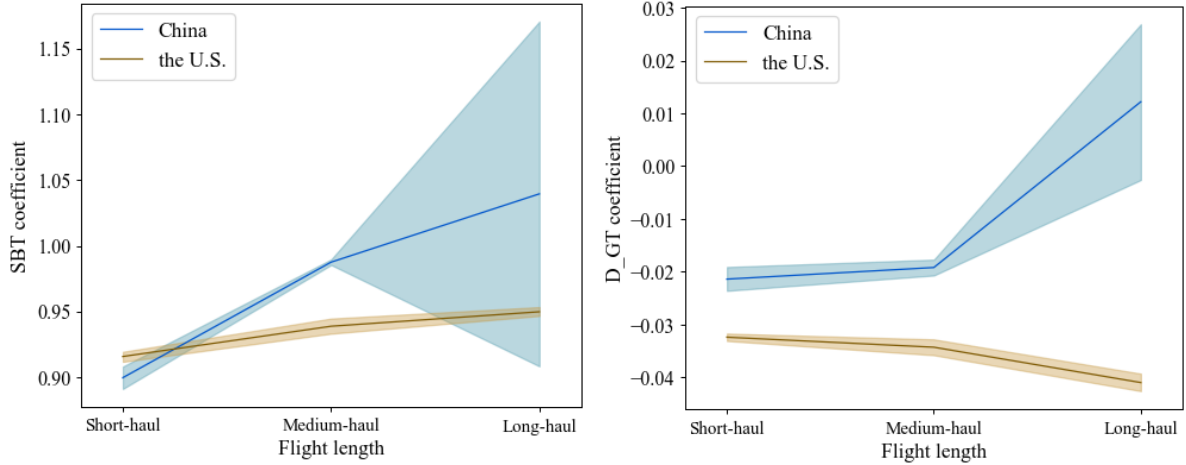

(a) SBT effect against NumFltOfAC     (b) D_GT effect against NumFltOfAC

Figure 4. Delay recovery effect variation with different flight length

Table 4. Estimation results of AAT for different flight length for China and the U.S.

|  | Short-haul | | Medium-haul | | Long-haul | |
| --- | --- | --- | --- | --- | --- | --- |
|  | China | U.S. | China | U.S. | China | U.S. |
| SBT | 0.900*** | 0.916*** | 0.988*** | 0.939*** | 1.040*** | 0.950*** |
| D_GT | -0.021*** | -0.032*** | -0.019*** | -0.034*** | 0.012 | -0.041*** |
| AF_TT_dif | 0.000*** | 0.001*** | 0.000*** | 0.000*** | 0 | 0.000*** |
| Heading | -0.166*** | 0.128*** | -0.018*** | 0.048*** | -0.406*** | 0.035*** |
| HubDep | -3.805*** | 0.410*** | -5.649*** | -0.315** | -3.386** | -2.145*** |
| HubArr | -0.798*** | -1.748*** | -2.210*** | -1.847*** | 1.027 | -3.611*** |
| Flt_last | -1.007*** | 0.318*** | -1.455*** | -0.001 | 0.779 | 0.417*** |
| DT | 1.022*** | -0.465*** | 0.143* | -0.679*** | -3.714** | -0.331*** |
| Wkd | 0.350*** | 0.078*** | 0.269*** | 0.472*** | 0.783*** | 0.467*** |
| Sprg | 1.200*** | 0.297*** | 2.317*** | 0.862*** | -1.265 | 0.758*** |
| Sumr | 0.865*** | -0.136*** | 1.115*** | 1.150*** | -0.672 | 0.994*** |
| Fall | 0.255** | -0.578*** | -1.056*** | -0.064 | -1.089 | 0.164 |
| VisDep | -0.403*** | 0.015 | 0.764*** | -0.028 | -1.325 | 0.463*** |
| TempDepDif | 0.029*** | -0.033*** | 0.077*** | -0.026*** | -0.116*** | 0.053*** |
| WdspDep | -0.073*** | -0.069*** | -0.111*** | -0.037*** | -0.121 | -0.103*** |
| VisArr | -0.221 | -1.387*** | -1.074*** | -2.034*** | -1.842*** | -1.812*** |
| TempArrDif | 0.007 | -0.013*** | 0.027*** | -0.021*** | 0.073*** | -0.033*** |
| WdspArr | 0.114*** | 0.058*** | 0.209*** | 0.054*** | 0.187** | 0.107*** |
| Cons | -11.279*** | -17.280*** | -22.770*** | -15.279*** | -23.78 | -15.623*** |
| N | 1,160,582 | 965,749 | 520,185 | 878,226 | 43,086 | 446,289 |
| R-sq | 0.856 | 0.87 | 0.836 | 0.87 | 0.836 | 0.918 |

*** $p<0.01$, ** $p<0.05$, * $p<0.10$.

## 5.2 Sensitivity to aircraft utilization

We measure aircraft utilization as the number of flight legs the aircraft associated with the flight under study flies in a day, and denote the number as NumFltOfAC. As mentioned in the beginning of Section 5, two opposing forces affect AAT as NumFltOfAC increases: 1) tendency to add flight buffer to mitigate delay propagation; 2) pressure to reduce flight buffer to fly more legs. The amount of flight buffer affects the difference between departure delay and arrival delay which in turn influences AAT (see Eq. (3)). Figure 5 illustrates the distribution of flights in seven groups based on NumFltOfAC: 1, 2, 3, 4, 5, 6, and >6. In China, about 70% flights are associated with aircraft with 4-6 legs per day, whereas in the U.S. the distribution is more skewed towards the lower end. This observation is consistent with the flight length distribution in the two countries: longer average OD distance per flight in the U.S. means that an aircraft can fly fewer legs in a day.

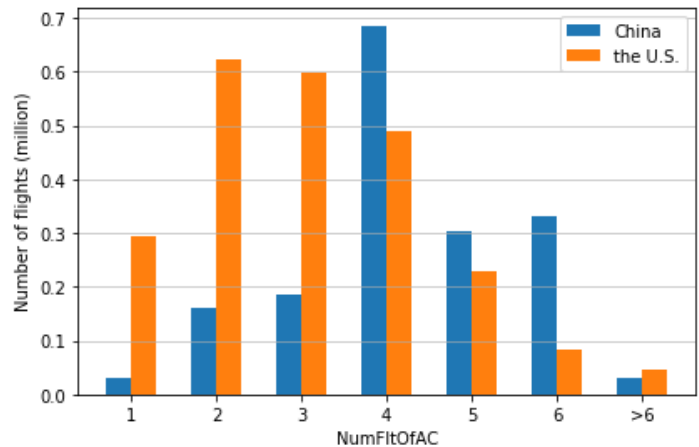

Figure 5. Distribution of the number of legs flown by the aircraft associated with the flights in China and the U.S.

Table 5 reports the GMM estimation results of the SBT model for each of the seven flight groups for each country. Again all standard errors are clustered by OD pair. Similar to reasongin in subsection 5.1, the focus of our analysis is on the coefficients for SBT and D_GT, for which point estimates and 95% confidence intervals are plotted in Figure 6. In Figure 6a, the SBT coefficient in China follows an overall decreasing trend as NumFltOfAC increases suggesting a dominating effect of adding flight buffer. The trend may be related to the heavy penalty for flight delays imposed by CAAC on Chinese airlines (CAAC, 2016). On the other hand, the changing trend for the U.S. flights is increasing, which means U.S. airlines are prone to reducing flight buffer towards tighter schedule to allow an aircraft to fly more legs in a day.

Figure 6b shows the D_GT coefficient estimates as a function of NumFltOfAC. Consistent with the base model estimates in Section 4, the ability of a flight to recover departure delay is limited in both countries. In China, the overall trend is a decrease in absolute value of the D_GT coefficient as NumFltOfAC increases (except when NumFltOfAC increases from one to two), whereas in the U.S. an overall increasing trend is observed. The difference is in line with the argument for interpreting the SBT coefficients. With more legs flown in a day, more buffer added to the flight schedule reduces the urgency and thus incentive for a Chinese flight to fly faster to make up for pre-takeoff time loss. By contrast, the tendency to have tigher schedule to accommodate more flight legs in a day in the U.S. puts more pressure on a flight to fly fast to stick to the schedule. But again, the effects are quite small.

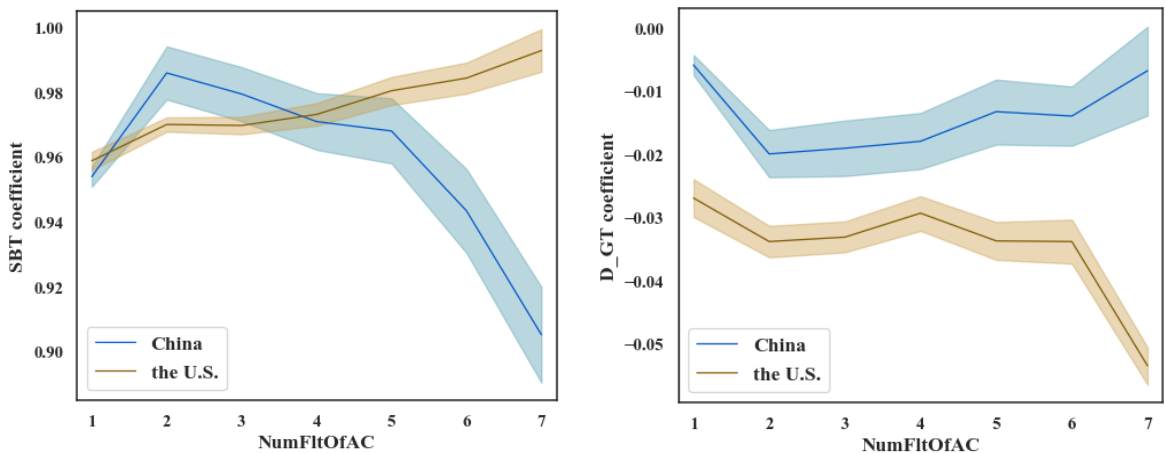

(a) SBT effect against NumFltOfAC  (b) D_GT effect against NumFltOfAC

Figure 6. Delay recovery effect variation with NumFltOfAC

Table 5. Estimation results of AAT in different NumFltOfAC of China and the U.S.

| China | 1 | 2 | 3 | 4 | 5 | 6 | >6 |
|---|---|---|---|---|---|---|---|
| SBT | 0.954*** | 0.986*** | 0.979*** | 0.971*** | 0.968*** | 0.943*** | 0.905*** |
| D_GT | -0.006*** | -0.020*** | -0.019*** | -0.018*** | -0.013*** | -0.014*** | -0.007* |
| AF_TT_dif | 5.04E-5*** | 7.06E-5*** | 7.94E-5*** | 9.86E-6 | -4.98E-5 | 7.05E-5*** | 3.50E-4*** |
| Heading | -0.139*** | -0.108*** | -0.097*** | -0.070** | -0.073** | -0.080*** | -0.027 |
| HubDep | -5.492*** | -4.911*** | -5.830*** | -4.203*** | -3.875*** | -4.313*** | -4.503*** |
| HubArr | -2.171*** | -1.057*** | -1.015*** | -0.965*** | -0.585* | -1.294*** | -1.571** |
| Flt_last | 0 | -0.841*** | 0.139 | -1.029*** | -0.991*** | -2.030*** | 0.151 |
| DT | 3.760*** | 1.681*** | 1.551*** | 0.515* | 0.565 | 0.25 | 0.217 |
| Wkd | 0.624*** | 0.329*** | 0.305*** | 0.284*** | 0.237*** | 0.359*** | 0.454*** |
| Sprg | 1.401*** | 2.453*** | 1.475*** | 1.440*** | 1.162*** | 1.174*** | 1.160*** |
| Sumr | 1.405*** | 1.674*** | 1.640*** | 0.899*** | 1.139*** | 0.346* | 0.593* |
| Fall | -0.656*** | 0.032 | -0.437** | -0.670*** | 0.093 | 0.008 | 0.007 |
| VisDep | 0.005 | -0.102 | -0.159 | 0.341** | 0.121 | -0.002 | 0.432** |
| TempDepDif | 0.069*** | 0.037*** | 0.060*** | 0.053*** | 0.032*** | 0.026*** | -0.027* |
| WdspDep | 0.015 | -0.064*** | -0.090*** | -0.071*** | -0.065** | -0.055** | -0.059* |
| VisArr | -0.291** | -1.286*** | -1.250*** | -0.959*** | -0.314 | -0.082 | 0.865*** |
| TempArrDif | 0.027*** | 0.023** | 0.028*** | 0.050*** | 0.033*** | 0.008 | 0.070*** |
| WdspArr | 0.145*** | 0.160*** | 0.206*** | 0.108*** | 0.135*** | 0.111*** | 0.136*** |
| Cons | -20.781*** | -22.821*** | -21.413*** | -19.269*** | -19.302*** | -14.557*** | -13.900*** |
| N | 29,813 | 159,612 | 184,616 | 684,834 | 304,639 | 329,959 | 30,380 |
| R-sq | 0.916 | 0.939 | 0.946 | 0.931 | 0.921 | 0.898 | 0.899 |

| U.S. | 1 | 2 | 3 | 4 | 5 | 6 | >6 |
|---|---|---|---|---|---|---|---|
| SBT | 0.961*** | 0.971*** | 0.970*** | 0.973*** | 0.981*** | 0.985*** | 0.994*** |
| D_GT | -0.019*** | -0.035*** | -0.033*** | -0.029*** | -0.034*** | -0.034*** | -0.055*** |
| AF_TT_dif | 2.82E-5*** | 3.53E-6 | 1.53E-5* | -3.7E-6 | 1.03E-4*** | 1.16E-4** | 6.90E-4*** |
| Heading | 0.047*** | 0.050*** | 0.050*** | 0.062*** | 0.061*** | 0.078*** | 0.075*** |
| HubDep | -1.830*** | -0.385** | -0.591*** | -0.575*** | -0.648*** | -0.751*** | 0.462* |
| HubArr | -2.829*** | -2.791*** | -2.861*** | -2.757*** | -2.855*** | -2.930*** | -2.184*** |
| Flt_last | 0 | -0.106 | -0.547*** | -1.044*** | -1.000*** | 0 | -2.355*** |
| DT | -0.124 | -0.621*** | -0.746*** | -0.908*** | -1.349*** | -1.185*** | -2.565*** |
| Wkd | 0.707*** | 0.403*** | 0.342*** | 0.284*** | 0.110* | 0.0764 | -2.648*** |
| Sprg | 1.251*** | 0.605*** | 0.653*** | 0.279** | 0.303** | 0.0856 | -1.512*** |
| Sumr | 1.430*** | 0.732*** | 0.717*** | 0.486*** | 0.456*** | 0.334* | -0.394*** |
| Fall | 0.072 | -0.180* | -0.144 | -0.388*** | -0.765*** | -0.417** | -0.990*** |
| VisDep | 0.468*** | 0.212*** | 0.047 | -0.011 | -0.168* | -0.161 | -0.823*** |
| TempDepDif | -0.020*** | -0.007 | -0.018*** | -0.026*** | -0.023*** | -0.019** | -0.068*** |
| WdspDep | -0.067*** | -0.061*** | -0.061*** | -0.041*** | -0.041*** | -0.043*** | -0.052*** |
| VisArr | -1.637*** | -1.963*** | -1.720*** | -1.725*** | -1.634*** | -1.084*** | -0.986*** |
| TempArrDif | -0.009 | -0.021*** | -0.027*** | -0.034*** | -0.035*** | -0.034*** | -0.068*** |
| WdspArr | 0.088*** | 0.076*** | 0.069*** | 0.069*** | 0.068*** | 0.092*** | 0.033*** |
| Cons | -20.67*** | -21.30*** | -20.81*** | -21.14*** | -21.47*** | -23.61*** | -18.62*** |

| N | 293,645 | 622,366 | 597,848 | 487,813 | 228,308 | 82,094 | 44,993 |
| --- | --- | --- | --- | --- | --- | --- | --- |
| R-sq | 0.977 | 0.983 | 0.982 | 0.977 | 0.975 | 0.976 | 0.950 |

*** $p<0.01$, ** $p<0.05$, * $p<0.10$.

# 6 Counterfactual analysis of potential efficiency gains in China

Given the difference in the behavior of AAT between China and the U.S., this section investigates how AAT would change and the implications for airline fuel consumption and operating cost if the AAT behavior in the U.S. were adopted in China. As discussed when introducing Hypothesis 4, the underlying conjecture is that due to more abundant airspace, flexible routing networks, and efficient ATFM procedures, adopting the U.S. AAT behavior would help reduce AAT and consequently airline fuel burn and operation cost in China.

## 6.1 Methodology

The basic idea for performing the counterfactual analysis is to apply the U.S. SBT model to the China data. As the air operation environment also affects SBT, it is important to account for the change in SBT. To do so, we run OLS regression of SBT on the instrumental variables using the U.S. data. The estimated model (termed "U.S. OLS model") is then applied to the China data to predict SBT. Note that the U.S. OLS model involves airline dummies and the Heading variable, for which applying the U.S. coefficients would be unrealistic. The inherent characteristics of China airlines would likely persist even if the air operation environment in China became the same as that of the U.S. Also, the Heading effect is location specific. Thus, when using the U.S. OLS model to predict SBT of the Chinese flights, the coefficients for the airline dummies and Heading still come from running the OLS regression of SBT using the China data (termed "China OLS model").

While most of the aircraft types appearing in the China dataset can be found in the U.S. flights, about 1% of Chinese flights have aircraft type that do not exist in the U.S. data. For these flights, their SBT assuming a U.S. environment is estimated using the China OLS model to first predict their SBT in the current environment, and then adding an average SBT change should the environment change.

For the average SBT change, we apply the U.S. OLS model and the China OLS model to all Chinese flights with common aircraft types in the two countries. The difference in the predicted SBT from the two models is averaged over all the flights to obtain the average SBT change.

Figure 7 illustrates the overall prediction procedure. For flights whose aircraft type can be found in both countries, the orange box indicates the use of the U.S. OLS model along with the airline dummy and Heading coefficients from the China OLS model, to generate predicted SBT (i.e., SBT′). For flights with aircraft types only in China but not in the U.S., the China OLS model is applied (blue box) with an adjustment by the average SBT change ($\bar{\Delta}_{SBT}$) to produce SBT′ of those flights. SBT′ is then used as an input to the U.S. SBT model (except for the Heading coefficient which comes from the China SBT model, as indicated by the grey box) and the China data to generate new AAT (i.e., AAT′).

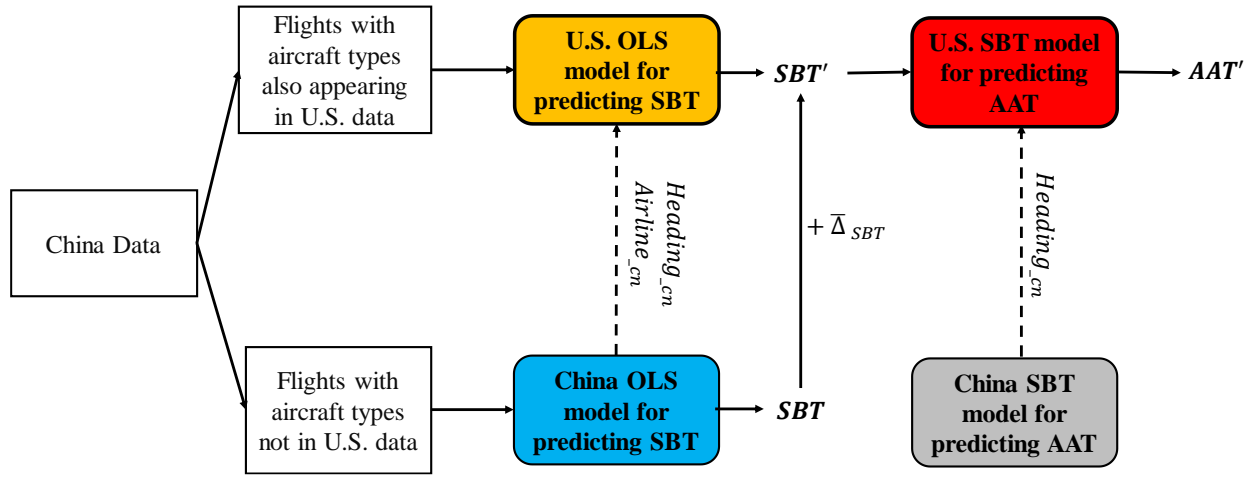

Figure 7. The procedure to predict AAT of Chinese flights if AAT behavior in the U.S. were adopted

## 6.2 Result

Figure 8 provides the distribution of AAT change over all Chinese flights if AAT behavior in the U.S. were adopted. Consistent with *a priori* expectation, most of the flights would experience a reduction in AAT. The average AAT change is a 11.8-minute reduction with a standard deviation of 8.6 minutes. Following the way AAT is predicted in the counterfactual, the change in AAT can be decomposed into two parts: one part is associated with SBT change; the other part is attributed to change in actual flying *conditional on* SBT. Our results show that the first part corresponds to an average of SBT reduction of 5.6 minutes among Chinese flights by adopting the AAt behavior in the U.S. The second part is the difference between the overall AAT reduction and the first part, which equals 6.2 minutes on average. Thus, AAT reduction would stem from both shorter SBT and faster flying relative to the new SBT. The second part is slightly larger than the first part.

The boxplots of the decomposed AAT change into SBT change and AAT change relative to SBT for the top 10 airlines in China (in terms of the number of flights) are shown in Figure 9. Except for Sichuan Airlines with a zero median in SBT change, all the other nine airlines would experience a reduction in SBT. Based on the median, we again observe that most of the airlines would experience a larger AAT reduction relative to the SBT than the reduction in SBT.

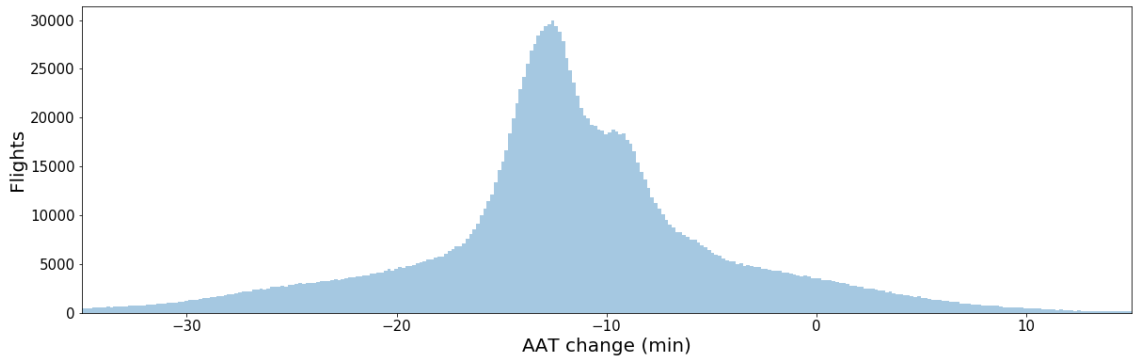

Figure 8. Distribution of AAT change among Chinese flights

if the AAT behavior in the U.S. were adopted

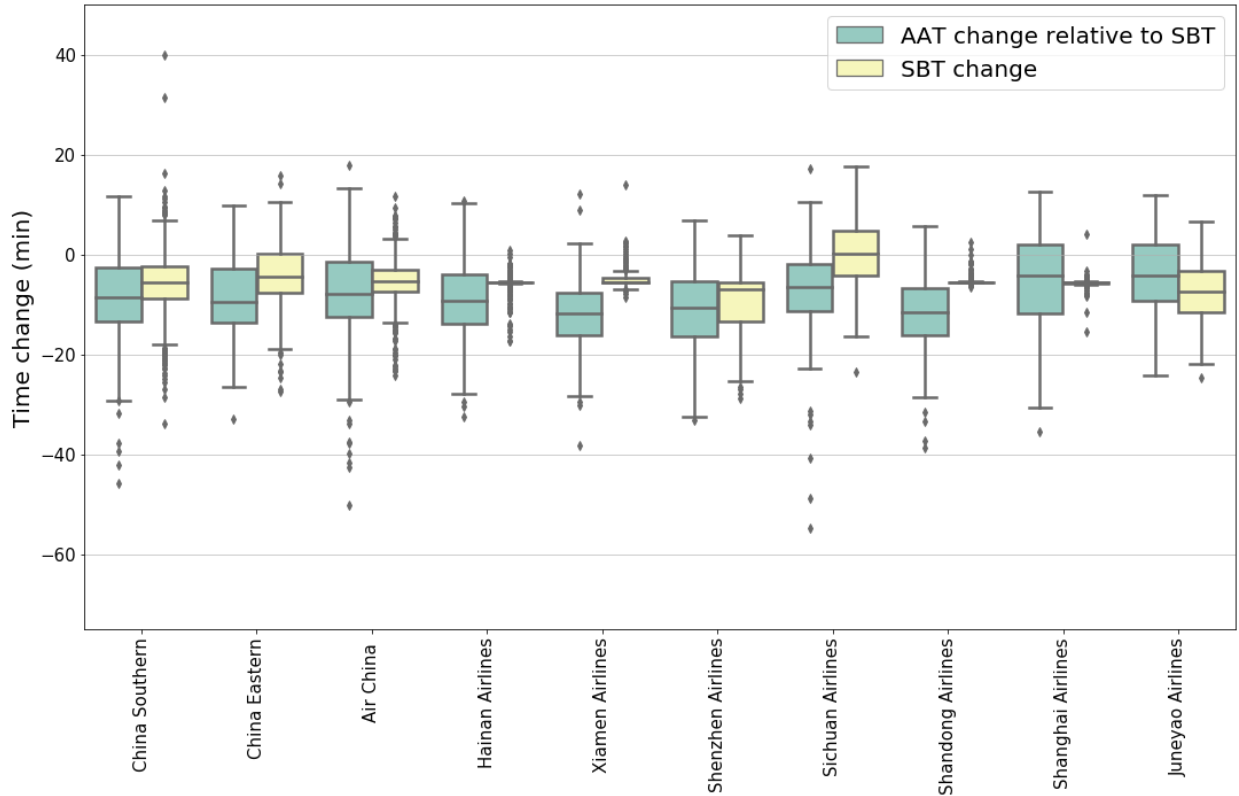

Figure 9. Boxplot of SBT change and AAT change relative to SBT for the top 10 Chinese airlines

The above estimates of AAT reduction can be used to assess the potential savings in airline fuel, fuel cost, and direct operating cost. Fuel consumption for each flight is calculated based on aircraft operating costs documentation of FAA (2016) and Statistical Data on Civil Aviation of China (2017). Table 7 reports the calculated fuel savings, per flight and in total for the four largest airlines (Air China, China Eastern, China Southern, and Hainan Airlines) and the rest combined. On average, 179.4 gallons of fuel would be saved per flight, with a systemwide saving of over 300 million gallons in 2016. The largest beneficiating airline would be China Southern which has the large number of flights in the dataset (over 20% in the total), whereas Air China would save the most on a per flight basis. Using the average fuel price of $1.45 per gallon in China in 2016, the fuel saving translates to $448.5 million cost savings.

Table 7. Estimates of time, fuel, fuel cost, and direct operating cost saving for Chinese airlines

| Airline | Time saving (min) | | Fuel saving (gallons) | | Fuel cost saving ($m) | Ratio | DOC saving ($m) |
|---|---|---|---|---|---|---|---|
| | Per flight | Total (million) | Per flight | Total (million) | | | |
| Air China | 13.0 | 2.5 | 229.2 | 43.3 | 62.8 | 0.331 | 189.9 |
| China Eastern | 10.9 | 2.8 | 164.6 | 42.5 | 61.7 | 0.340 | 181.4 |
| China Southern | 13.0 | 2.5 | 204.2 | 71.2 | 103.3 | 0.370 | 279.0 |
| Hainan Airline | 12.5 | 1.7 | 212.7 | 29.4 | 42.6 | 0.346 | 132.4 |
| Minor Airlines (184 airlines) | 11.3 | 10.5 | 164.1 | 152.2 | 220.7 | 0.359 | 510.0 |
| **Total** | **11.8** | **20.3** | **179.4** | **309.3** | **448.5** | / | **1,283.7** |

Notes:
1. The ratio for the minor airlines is derived from the average value of Juneyao Airlines and Spring Airlines.
2. As the major airlines hold direct or indirect interests in their subsidiary airlines, the same ratio of fuel cost to direct operating cost is adopted. Below is the correspondence between major and subsidiary airlines:
    a) Air China: Shenzhen Airline, Kunming Airline, Air Macau, Dalian Airlines and Air China Inner Mongolia, Shandong Airline, Tibet Airlines;
    b) China Eastern: Shanghai Airlines, China United Airlines;
    c) China Southern: Xiamen Airlines, Chongqing Airlines, Sichuan Airlines;
    d) Hainan Airline: Air Chang'an, Beijing Capital Airlines, Fuzhou Airlines, Grand China Air, Lucky Air, Tianjin Airlines, Urumqi Airlines.

The reduction in AAT saves not only fuel consumption but also other cost related to flight operations, to estimate which we multiply fuel cost saving by the ratio of fuel cost in total direct operating cost of the airlines. The total direct operating cost consists of fuel cost, aircraft depreciation, crew salaries, and aircraft maintenance and rentals. For Air China, China Eastern, China Southern, and Hainan Airlines, the relevant cost information can be obtained from their annual income statements (Air China LLC., 2016; China Eastern LLC., 2016; China Southern Airlines LLC., 2016; Hainan Airlines LLC., 2016). For the other airlines, the ratio is based on the average of the ratios of Juneyao Airlines and Spring Airlines, the only two airlines for which cost information is available to us (Juneyao Airlines LLC., 2016; Spring Airlines LLC., 2016). As shown in Table 7, the ratios are quite consistent across airlines, in the range of 0.33-0.37. The resulting total direct operating cost saving amounts to nearly $1.3 billion, again with the largest saving reaped by China Southern, close to $300 million in 2016. Figure 10 shows the top 10 airlines in terms of the total direct operating cost saving. Note that Shenzhen Airlines, Xiamen Airlines, Shanghai Airlines, Shandong Airlines, and Sichuan

Airlines are subsidiaries of Air China, China Eastern, and China Southern. Finally, recall that the data are filtered and do not include all flights. The original data contain 3,135,075 domestic commercial flights. As a first-order extrapolation, the nationwide saving of total direct operating cost would amount to about $2.3 billion for year 2016.

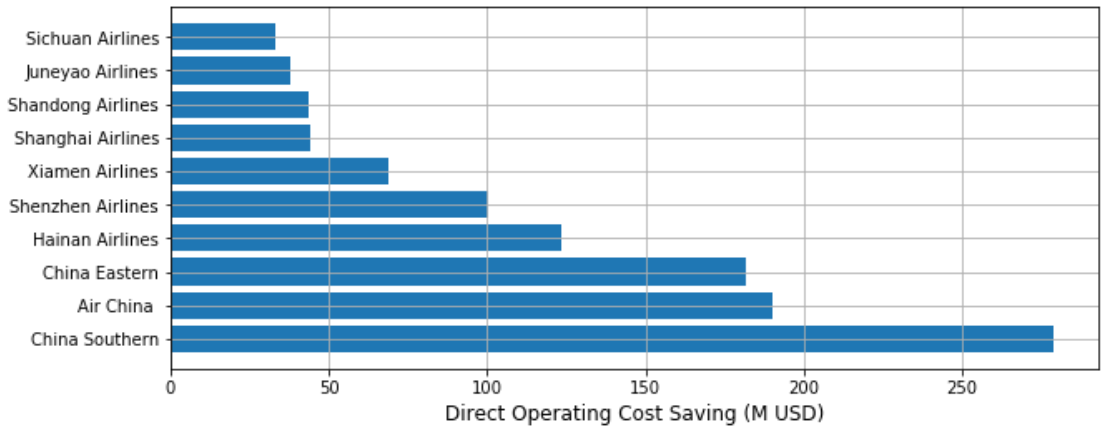

Figure 10. Direct operating cost saving for the top 10 airlines in China

We further compute the average AAT change for routes that had at least 60 flights in 2016. Routes with too few flights may imply that flights were scheduled on a more *ad-hoc* basis, thus less representative for the purpose of route-level analysis. Figure 11 visualizes the average AAT change by route. The lines in Figure 11a indicate routes with a decrease in AAT, whereas the lines in Figure 11b depict the routes with an increase in AAT. The thickness of each line corresponds to the extent of AAT reduction. In total, 581 routes would experience reduced AAT, with the largest amount of reduction occurring to Shenyang (SHE)-Sanya (SYX) (67 minutes), Changchun (CGQ)-Shenzhen (SZX) (60 minutes), and Shenyang (SHE)-Shenzhen (SZX) (60 minutes). These routes are among the longest north-south routes in eastern part of China. In contrast, 137 routes would experience increased AAT, with the largest amounts occurring to Ürümqi (URC)-Shanghai Pudong (PVG) (37 minutes), Ürümqi (URC)-Hohhot (HET) (31 minutes), and Ürümqi (URC)-Tianjin (TSN) (31 minutes).

The routes with increased and decreased AAT indeed follow distinctive spatial patterns. As shown in Figure 11, the routes with increased AAT are most north-south bound, whereas the routes with decreased AAT are predominately east-west bound. Such spatial patterns can be explained by the oppositie concentration of air traffic in the two countries. Recall from Figure 2 that most of the routes among the top 30 airports in China are north-south bound, whereas most of the routes among the top 30 airports in the U.S. are east-west bound. In the U.S., the higher air traffic volume for east-west bound than for north-south bound means north-south bound flying are more efficient than east-west bound flying. By adopting the AAT behavior in the U.S., the north-south routes in China, which accommodate majority of air traffic in the country and thus suffer more congestion than flights on other routes, would enjoy improved flying efficiency. On the other hand, the east-west routes in China accommodate less traffic resulting in relatively uncongested flying in the current environment. Adopting the AAT behavior in the U.S., which represents less efficient flying on the east-west routes, would result in lowered efficiency of flying on these east-west routes in China. In summary, the counterfactual would introduce more efficient flying to most routes with heavy traffic, leading to a total reduction of AAT despite some routes experiencing AAT increase. An overall net gain will accrue to the Chinese air traffic system.

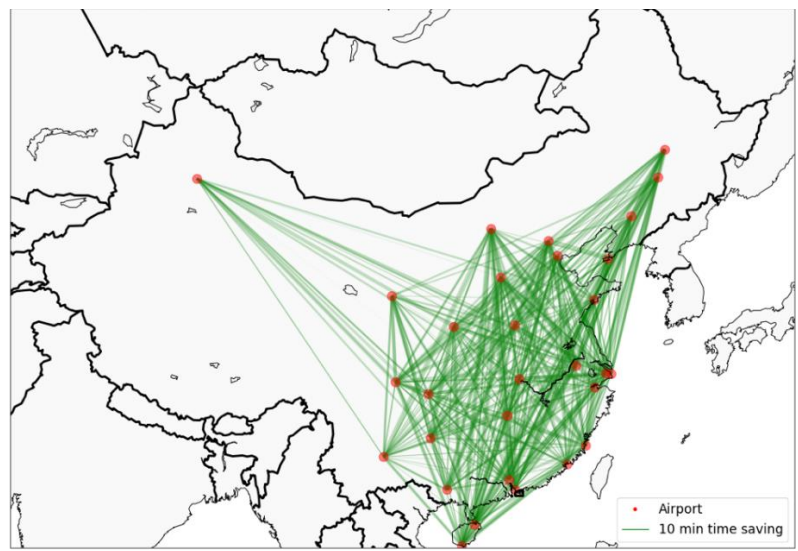

(a) Routes with decreased AAT

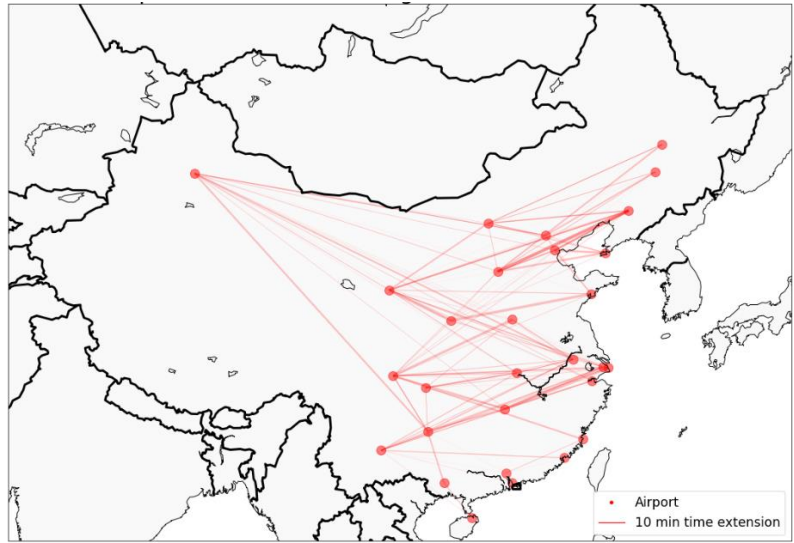

(b) Routes with increased AAT

Figure 11. AAT change among Chinese routes if the AAT behavior in the U.S. were adopted

# 7 Conclusion

Understanding the behavior of flight actual airbone time has become increasingly important given the ever growing air travel demand and flight delays becoming more rampant. For the first time in the literature, this paper performs an empirical investigation of AAT and its influencing factors, particularly how AAT is affected by SBT, OD distance, and the possible pressure to reduce AAT from other parts of flight operations. A uniqueness of our investigation is its comparative nature: the modeling of AAT is side-by-side for the U.S. and China, which have the two largest air traffic systems in the world with distinct characteristics in terms of geographical distributions of air travel demand, airspace availability, air traffic flow management, delay management policies by civil aviation authorities, and airline scheduling practices.

The different model specifications and empirical estimation yield a number of interesting insights, with the most important ones reflected in the testing of the four hypotheses. Specifically, AAT is highly correlated with SBT and OD distance. Flights in the U.S. are faster on average than in China due to more abundant airspace and consequently greater possibility to fly great circle distance. Facing ground delay prior to takeoff, a flight has limited capability to speed up. The pressure from short turnaround time after landing to reduce AAT is found immaterial. These findings suggest that the established ATC procedures and airspace congestion do not allow much flexibility in AAT for the purpose of mitigating delays. The findings are consistent between the two countries. Sensitivity analysis of AAT to flight length and aircraft utilization is further conducted. The results show some discrepancy in the changing trend of AAT in China and the U.S., reflecting the underlying differences in the operation environment in the two countries. Finally, given the more abundant airspace, flexible routing networks, and efficient ATFM procedures, a counterfactual that the AAT behavior of the U.S. were adopted in China is examined which suggests that significant efficiency gains would be achieved in terms of AAT, fuel use, and direct airline operating cost.

This research presents a beginning towards understanding the behavior of actual flight airborne time. Future research can be extended in a few directions. First, as en-route traffic and weather conditions, and ATC all affect AAT behavior, collecting and incorporating relevant information may help further enhance model fit and interpretability. Second, provided that data become available in China, we could decompose D_GT into gate departure delay and taxi-out delay, and explore whether the two types of pre-takeoff delays have the same influence on AAT. Third, as flights flown by the same aircraft in a day are connected, it would be interesting to explore modeling AATs for a series of flight legs to further understand the behavior of AAT of not just a single flight but multiple ones and their interdependence.

# Appendix 1: Major hub airports in China and the U.S.

|  | Airlines | Hub airports |
|---|---|---|
| China | Air China | Beijing Capital International Airport (PEK)<br>Chengdu Shuangliu International Airport (CTU)<br>Shanghai Pudong International Airport (PVG)<br>Shanghai Hongqiao International Airport (SHA) |
| | Eastern Airlines | Shanghai Pudong International Airport (PVG)<br>Shanghai Hongqiao International Airport (SHA)<br>Beijing Capital International Airport (PEK)<br>Kunming Changshui International Airport<br>Xi'an Xianyang International Airport (XIY) |
| | Southern Airlines | Guangzhou Baiyun International Airport (CAN)<br>Beijing Capital International Airport (PEK)<br>Ürümqi Diwopu International Airport (URC)<br>Chongqing Jiangbei International Airport (CKG)<br>Shanghai Pudong International Airport (PVG) |
| | Hainan Airlines | Haikou Meilan International Airport (HAK)<br>Beijing Capital International Airport (PEK)<br>Xi'an Xianyang International Airport (XIY) |
| | Xiamen Airlines | Xiamen Gaoqi International Airport (XIA)<br>Fuzhou Changle International Airport (FOC)<br>Hangzhou Xiaoshan International Airport (HGH)<br>Changsha Huanghua International Airport (CSX) |
| | Shanghai Airlines | Shanghai Pudong International Airport (PVG)<br>Shanghai Hongqiao International Airport (SHA) |
| | Shandong Airlines | Jinan Yaoqiang International Airport (TNA)<br>Qingdao Liuting International Airport (TAO)<br>Xiamen Gaoqi International Airport (XIA) |
| U.S. | Southwest Airlines | Denver International Airport (DEN)<br>McCarran International Airport (LAS)<br>Phoenix Sky Harbor International Airport (PHX)<br>Hartsfield-Jackson Atlanta International Airport (ATL)<br>Orlando International Airport (MCO)<br>Chicago Midway Airport (MDW)<br>Baltimore/Washington International Thurgood Marshall Airport (BWI) |
| | United Airlines | San Francisco International Airport (SFO)<br>Los Angeles International Airport (LAX)<br>Denver International Airport (DEN)<br>George Bush Intercontinental Airport (IAH)<br>Newark Liberty International Airport (EWR)<br>Chicago O'Hare International Airport (ORD)<br>Washington Dulles International Airport (IAD) |
| | Delta Airlines | Hartsfield-Jackson Atlanta International Airport (ATL)<br>Detroit Metropolitan Wayne County Airport (DTW)<br>Minneapolis-Saint Paul International Airport (MSP)<br>Los Angeles International Airport (LAX) |

| | | Salt Lake City International Airport (SLC) |
| | | John F. Kennedy International Airport (JFK) |
| | American Airlines | Philadelphia International Airport (PHL) |
| | | Phoenix Sky Harbor International Airport (PHX) |
| | | Dallas-Fort Worth International Airport (DFW) |
| | | John F. Kennedy International Airport (JFK) |
| | | Los Angeles International Airport (LAX) |
| | | Chicago O'Hare International Airport (ORD) |
| | | Charlotte Douglas International Airport (CLT) |

# Appendix 2: Components in the direct operation cost of major and some minor Chinese airlines

| 2016 (billion RMB) | Fuel | Depreciation | Crew salaries | Aircraft Maintenance | Rentals | Total | **Ratio** |
|---|---|---|---|---|---|---|---|
| **Major Airlines** | | | | | | | |
| Air China | 21.98 | 13.47 | 20.08 | 4.65 | 6.25 | 66.43 | **0.33** |
| China Eastern | 19.63 | 12.15 | 18.15 | 4.96 | 2.86 | 57.75 | **0.34** |
| China Southern | 23.80 | 12.62 | 92.15 | 11.32 | 7.33 | 64.28 | **0.37** |
| Hainan Airline | 7.86 | 3.66 | 2.88 | 3.99 | 4.35 | 22.74 | **0.35** |
| **Minor Airlines** | | | | | | | |
| Juneyao Airlines | 1.92 | 1.53 | 1.57 | 0.44 | 0.14 | 5.61 | **0.34** |
| Spring Airlines | 1.88 | 1.33 | 1.39 | 0.41 | / | 5.01 | **0.38** |